\documentclass[aps,prl,twocolumn,groupedaddress]{revtex4-1}
\usepackage[dvips]{graphicx}
\usepackage{bm}
\usepackage{color}
\usepackage{mathtools}

\begin{document}
\title{Quantum Mechanics with a Momentum-Space Artificial Magnetic Field}

\author{Hannah M. Price, Tomoki Ozawa and Iacopo Carusotto}

\affiliation{INO-CNR BEC Center and Dipartimento di Fisica, Universit\`{a} di Trento, I-38123 Povo, Italy} 

\bigskip

\bigskip

\begin{abstract}
The Berry curvature is a geometrical property of an energy band which acts as a momentum space magnetic field in the effective Hamiltonian describing single-particle quantum dynamics. We show how this perspective may be exploited to study systems directly relevant to ultracold gases and photonics.  Given the exchanged roles of momentum and position, we demonstrate that the global topology of momentum space is crucially important. We propose an experiment to study the Harper-Hofstadter Hamiltonian with a harmonic trap that will illustrate the advantages of this approach and that will also constitute the first realization of magnetism on a torus.  
\end{abstract}

\maketitle

The Hamiltonian of a charged particle in an electromagnetic field is a familiar and fundamental result in quantum mechanics\cite{Landau1981Quantum}. In this Hamiltonian,
\begin{equation}
\mathcal{H}=\frac{({\bf p}-e {\bf A}({\bf r}))^2}{2M}+e\Phi({\bf r}) , ¤
\label{eq:magh}
\end{equation}
the roles of momentum and position are inherently asymmetric; the magnetic vector potential, ${\bf A}({\bf r})$, is a function of position which redefines the relationship between the canonical, ${\bf p}$, and physical, ${\bf p}-e{\bf A}({\bf r})$, momenta. The vector potential ${\bf A}({\bf r})$ is also responsible for the geometric Aharanov-Bohm phase, which depends on the real-space trajectory of a particle.

The magnetic Hamiltonian has an important {\it momentum space} counterpart,
\begin{equation}
\tilde{\mathcal{H}}=E({\bf p})+W({\bf r}+ \boldsymbol{\mathcal{A}}({\bf p})),
\label{eq:key}
\end{equation}
that underlies many intriguing phenomena in solid state physics such as the anomalous\cite{adams1959energy, nagaosa, haldanefs} and spin Hall effects\cite{murakami2003dissipationless, bliokh2005spin, fujita} as well as peculiar features of graphene\cite{novonobel, zhangnature} and bulk Rashba semiconductors\cite{Murakawa}.
In this formalism, $E({\bf p})$ is the energy dispersion of the band under consideration, while $\boldsymbol{\mathcal{A}}({\bf p})$ is the geometrical Berry connection of the band (defined below) \cite{adams1959energy, bliokh2005spin, berry}. The Berry connection acts as a momentum space vector potential, redefining the relationship between the canonical, ${\bf r}$, and physical, ${\bf r}+\boldsymbol{\mathcal{A}}({\bf p})$, position operators appearing in the external potential term $W({\bf r}+\boldsymbol{\mathcal{A}}({\bf p}))$. This replacement has important physical consequences that have been studied primarily, so far, at the semiclassical level \cite{adams1959energy, nagaosa, murakami2003dissipationless, bliokh2005spin, fujita, bliokh2005topological,gosselin2006semiclassical}. As in the Aharanov-Bohm effect, a particle moving in momentum space under the influence of an external force gains a geometrical Berry phase due to the connection $\boldsymbol{\mathcal{A}}({\bf p})$. The curvature, ${\bf \Omega}({\bf p})=\nabla_{\bf p}\times \boldsymbol{\mathcal{A}}({\bf p})$, also naturally defines a momentum space magnetic field\cite{berry, bliokh2005spin, PhysRevD.12.3845, cooper2012designing}.

Local geometrical properties of energy bands can be related to global topological invariants. For example, the simplest topological invariant of a 2D crystal, the so-called Chern number $\mathcal{C}$, is the integral of the Berry curvature over the first Brillouin zone (BZ). In the analogy with magnetism, the Chern number is the momentum space counterpart of the number of magnetic monopoles\cite{Fang} inside a torus. This important invariant underlies the quantization of conductance in the quantum Hall effect\cite{thouless}, while other topological invariants can be defined to classify topological insulators\cite{hasankane, qizhang}. 

In the last few years, geometrically nontrivial bands have been created in ultracold gases\cite{esslinger,strucksengstock,aidelsburger, hiro} and photonic systems\cite{wang2009, rechtsman2013, hafezi2, jacqmin}. nonzero ${\bf \Omega}({\bf p})$ can have consequences for the collective modes of an ultracold atomic gas\cite{duine, pricemodes} and for the semiclassical dynamics of a wave packet\cite{dudarev,1chang, price, cominotti, dauphin, tomoki}, while the hallmarks of nontrivial topological bands have been observed in topologically protected photonic edge states\cite{wang2009, rechtsman2013, hafezi2}.

In this Letter, we discuss how the momentum space magnetic Hamiltonian (\ref{eq:key}) can be exploited as a fully quantum theory to understand the quantum mechanics of single particles in energy bands with nontrivial geometrical and topological properties, in the presence of additional external potentials. 
To illustrate this most clearly, we focus on the example of a two-dimensional system where the energy and the Berry curvature of the lowest band are nearly flat over the first BZ. In the presence of an external harmonic potential, the equispaced eigenstates are then the momentum space counterpart of Landau levels in a constant magnetic field. Remarkably, these eigenstates have novel features directly stemming from the global toroidal topology of the BZ. The recent experimental realizations of the Harper-Hofstadter model in ultracold gases\cite{aidelsburger, hiro}, photonic systems\cite{hafezi2} and solid-state superlattices\cite{dean2013hofstadter} suggest a prompt experimental implementation of our approach. This would open up new avenues to experimentally investigate quantum mechanics and quantum magnetism on a topologically nontrivial manifold such as a torus\cite{jain, AlHashimi}.

{\it The effective quantum Hamiltonian.}-- We start by presenting a short derivation of the momentum space magnetic Hamiltonian (\ref{eq:key}) in modern terms for systems of current experimental interest. Our derivation builds on ideas over the last 60 years\cite{karplus, adams1959energy, berry} and is applicable to the generic single-particle Hamiltonian, $\mathcal{H} = \mathcal{H}_0+ W(\mathbf{r})$, whose first term, $\mathcal{H}_0$, is either translationally invariant or periodic in real space. For example, $\mathcal{H}_0$ could refer to an electron in a crystal, an atom with spin-orbit coupling, an ultracold atomic gas in an optical lattice or light in either a photonic crystal or a lattice of coupled resonators or waveguides. The second part of the Hamiltonian, $W({\bf r})$, is a weaker additional potential. This could be, for instance, an external static electric field for an electron, a harmonic trap or optical superlattice potential for atoms, or a slow modulation of the background refractive index and/or of the cavity size in optical systems.

The eigenfunctions of $\mathcal{H}_0$ are $| \chi_{n, {\bf p}}  ({\bf r}) \rangle= \frac{e^{ i {\bm p} \cdot {\bf r}}}{ \sqrt{V} }| n {\bf p}\rangle$, where $| n {\bf p}\rangle$ is the energy eigenstate for band index $n$ and momentum ${\bf p}$, and $V$ is a normalization factor. If $\mathcal{H}_0$ is periodic, the eigenstate is the periodic Bloch function, $u_{n, \mathbf{p}} ({\bf r})$, and the momentum is the crystal momentum defined in the BZ (we take $\hbar=1$ throughout). The normalization, $V$, is the number of lattice sites, $N$. If instead $\mathcal{H}_0$ is translationally invariant, the eigenstate $| n {\bf p}\rangle$ is independent of position and $V$ is the volume of the system. For simplicity, we focus on two dimensions, although the extension to 3D is straightforward. 

The energy bands have a band structure, $E_n (\mathbf{p})$, and geometrical properties encoded in the Berry connection, $\boldsymbol{\mathcal{A}}_{n} ({\bf p})$, and Berry curvature, $\Omega_{n}({\bf p})$\cite{berry,di}:
\begin{eqnarray} 
\boldsymbol{\mathcal{ A}}_{n} ({\bf p}) & \equiv & i \langle n {\bf p}| \frac{\partial}{\partial {\bf p}} |n {\bf p}\rangle , \hspace{0.2in}
\Omega_{n}({\bf p}) \equiv  {\bm \nabla_{\bf p}} \times \boldsymbol{\mathcal{ A}}_{n} ({\bf p}) \cdot\hat{{\bf z}}. \nonumber
\end{eqnarray}  
The additional potential, $W({\bf r})$, mixes different eigenstates, $|n {\bf p}\rangle$. We expand the eigenstates of the full Hamiltonian, $\mathcal{H}$, as $
	|\Psi\rangle = \sum_{n}\sum_{\bf p} \psi_{n } (\mathbf{p})  |\chi_{n,\mathbf{p}}\rangle$,
where $\psi_{n }(\mathbf{p})$ are expansion coefficients. For a periodic $\mathcal{H}_0$, this sum is taken over the first BZ, otherwise, the sum runs over all momenta. We substitute into the Schr\"odinger equation, $i\frac{\partial}{\partial t}|\Psi\rangle=\mathcal{H}|\Psi\rangle$, and apply $\langle \chi_{ n^\prime, \mathbf{p}^\prime}|$, to obtain:
\begin{eqnarray}
	i\frac{\partial}{\partial t}\psi_{n} (\mathbf{p}) 
	=
	E_n (\mathbf{p}) \psi_{n}(\mathbf{p}) +	 \sum_{n, \bf p} \langle \chi_{n^\prime,\mathbf{p}^\prime}|  W({\bf r}) |\chi_{n,\mathbf{p}}\rangle \psi_{n} (\mathbf{p}).  \nonumber \label{eq:sc2}
\end{eqnarray} 
We expand $W({\bf r})$ as a power series in ${\bf r}$, and repeatedly insert the completeness relation: $1=  \sum_n  \sum_{\bf p}  |\chi_{n,{\bf p}}\rangle \langle \chi_{n,{\bf p}}| $ (demonstrated explicitly in the Supplemental Material for a harmonic trap). Then we can use the identity:
\begin{align}
\langle \chi_{n^\prime,\mathbf{p}^\prime}|\mathbf{r}|\chi_{n, \mathbf{p}}\rangle
	&=
		\delta_{\mathbf{p},\mathbf{p}^\prime}
	\left(\delta_{n,n^\prime} i\nabla_{\mathbf{p}}	 +
	 i  \langle n^\prime\mathbf{p}^\prime| \frac{\partial}{\partial {\bf p}} | n{\bf p}\rangle\right) , \nonumber
\end{align}	
which we have generalized from a previously known result\cite{karplus, adams1959energy}. We assume that the additional potential is sufficiently weak that it does not significantly mix energy bands and that the contribution from only one band $n$ is non-negligible. A quantitative condition for this approximation will be discussed in the following. The effective quantum Hamiltonian in the single-band approximation then has the form (\ref{eq:key}) with the suitable $E_n({\bf p})$ and $\boldsymbol{\mathcal{A}}_n({\bf p})$. Of course, this Hamiltonian may be generalized to systems with degeneracies such as graphene and topological insulators\cite{hasankane, qizhang}; then, the effective momentum space magnetic field has a non-Abelian gauge structure\cite{bliokh2005spin, wilczek}.

{\it Connections with magnetism.}-- 
The duality between momentum space magnetism and real space magnetism is transparently demonstrated by comparing the effective Hamiltonian (\ref{eq:key}) to the textbook magnetic Hamiltonian (\ref{eq:magh}). The energy bandstructure, $E_n (\mathbf{p})$, acts like the external scalar potential $e\Phi({\bf r})$, while the external potential $W \left( i {\nabla_{\mathbf{p}}}+\boldsymbol{\mathcal{A}}_{n}(\mathbf{p})\right)$ corresponds to the ``kinetic energy", $\frac{1}{2M}{({\bf p}-e {\bf A}({\bf r}))^2}$~\footnote{Previously discussed specifically for a particle with 2D Rashba spin-orbit coupling in a harmonic trap in Ref.~\onlinecite{rashbalandau}}. For a harmonic trapping potential, $W ({\bf r})= \frac{1}{2} \kappa {\bf r}^2$, the effective Hamiltonian (\ref{eq:key}) is:
\begin{eqnarray}
	\tilde{\mathcal{H} }	&=& E_n (\mathbf{ p}) + 
	  \frac{\kappa
	\left(i {\bm \nabla}_{\mathbf{p}}
	+
	\boldsymbol{\mathcal{A}}_{n}(\mathbf{p})
	\right)^2}{2}, \label{eq:berry}
\end{eqnarray} 
where the inverse trapping strength, $\kappa^{-1}$ acts as the particle mass, $M$. (This is further illustrated in the Supplemental Material, for the toy model of a harmonic trap in an optical lattice without a momentum space magnetic field.) We focus hereafter on $W ({\bf r})= \frac{1}{2} \kappa {\bf r}^2$, but other forms of the energy-momentum relationship in the real space magnetic Hamiltonian could be obtained by applying different types of external potential, $W({\bf r})$. 

{\it The topology of momentum space.}-- The global properties of the Berry connection and curvature have been deeply investigated as they are related to topological invariants, underlying, for example, the quantum Hall effect~\cite{thouless}. However, much less attention has been devoted to the impact of the global topology of momentum space on the particle wave function which must be single valued~\cite{Landau1981Quantum}. This condition is irrelevant when the momentum ${\bf p}$ can take arbitrarily large values, e.g. for particles with 2D Rashba spin-orbit coupling in a Zeeman field~\cite{pricemodes}, but has very interesting consequences in spatially periodic systems where the momentum is defined over the BZ, which has the topology of a torus. As a concrete example of this, we investigate the Harper-Hofstadter Hamiltonian~\cite{hofstadter} with an external harmonic trap; this is a natural extension of recent experimental advances~\cite{hafezi2,aidelsburger, hiro, dean2013hofstadter}. 

{\it The Harper-Hofstadter model.}-- In the Harper-Hofstadter model, a particle hops on a 2D lattice in a perpendicular (real or artificial) magnetic field, ${\bm B} = B \hat{{\bf z}}$. In the Landau gauge, ${\bf A } ({\bf r}) = B x \hat{{\bf y}}$, the tight-binding Hamiltonian with a harmonic trap is:
\begin{eqnarray}
\hspace{-0.25in}{\mathcal{H}} &=& \mathcal{H}_0+ \frac{1}{2} \kappa a^2 \sum_{m, n}  (m^2+n^2) \hat{a}^\dagger_{m,n} \hat{a}_{m,n} ,  \nonumber \\ 
\hspace{-0.25in}\mathcal{H}_0 &=& - J \sum_{m, n}  \left( \hat{a}^\dagger_{m+1,n} \hat{a}_{m,n}+ e^{i \phi }  \hat{a}^\dagger_{m, n+1} \hat{a}_{m,n} \right) + \mbox{h.c.}\label{eq:hof}
\end{eqnarray}
where $\mathcal{H}_0$ is the Harper-Hofstadter Hamiltonian, $J$ is the hopping amplitude, $a$ is the lattice spacing and the $ \hat{a}_{m,n}^\dagger$ ($\hat{a}_{m,n}$) operators create (annihilate) a particle at lattice site $(m, n)$. The hopping along $\hat{{\bf y}}$ is modified by a complex phase $\phi = 2 \pi \alpha  m a$, where $\alpha$ is the number of magnetic flux quanta per plaquette of the lattice.

Without a harmonic trap, the eigenstates are those of the Harper-Hofstadter Hamiltonian, with behavior governed by the value of $\alpha$. When $\alpha = p/q$, the tight-binding band splits into $q$ magnetic subbands. The energy spectrum is the well-known Hofstadter butterfly\cite{hofstadter}. The magnetic vector potential, ${\bf A } ({\bf r})$, is not periodic, and the usual translation operators do not commute with $\mathcal{H}_0$\cite{ 1chang}. To apply Bloch's theorem, we define new magnetic translation operators and a larger magnetic unit cell of $q$ plaquettes, that contains an integer number of magnetic flux quanta. The Bloch states are then {\it magnetic} Bloch states defined within the {\it magnetic} Brillouin zone (MBZ): $-\pi/a <  p_y \leq \pi /a$ and $-\pi/ q a <  p_x \leq \pi /q a$ (for a magnetic unit cell of $q$ plaquettes along $\hat{{\bf x}}$) \cite{ 1chang}. 

 \begin{figure}[!]
  \centering
    $
  \begin{array}{c}
\resizebox{0.45\textwidth}{!}{\includegraphics*{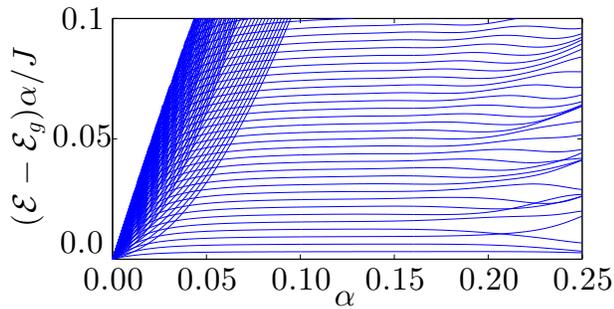}  }
\end{array}$
\caption{The numerical energy spacing relative to the lowest eigenvalue, $\mathcal{E}_g$, for the lowest 100 states, obtained by diagonalizing Eq. 4 for $\kappa a^2 / J = 0.02$ and a lattice with $N= 41 \times 41$ sites. The energy spacing is multiplied by $\alpha$ to highlight the spacing of toroidal Landau levels in the lowest band.} \label{fig:energyspacing} 
\end{figure}
{\it Numerical calculations with a harmonic trap.}-- Adding a harmonic trap splits the Harper-Hofstadter bands into a complicated structure first noted in Ref.~\onlinecite{kolovskypara} and replotted here in Fig. \ref{fig:energyspacing} in terms of the energy level spacing relative to the lowest numerical eigenstate. For each value of $\alpha$, the spectrum was obtained by numerically diagonalizing the full Hamiltonian (\ref{eq:hof}). The only significant error is from the restriction of (\ref{eq:hof}) to a finite lattice. To control this, we ensure that all energies are converged to within the accuracy shown. Numerical diagonalization also gives the real space eigenstates of (\ref{eq:hof}); we relate these to the population in the MBZ via a procedure described in the Supplemental Material. 

{\it Analytical interpretation.}--To understand the complicated spectrum, we build a simple model, focusing on $\alpha = 1/q \ll 1$ where our interpretation is the clearest. In this regime, we can make two simplifications; firstly, with decreasing $\alpha =1/q$, the bands flatten compared to the hopping energy $J$. If the bandwidth is much smaller than the harmonic trapping energy, we can assume $E_n ({\bf p}) \simeq E_n$, contributing only an overall energy shift. Secondly, when $\alpha=1/q$ and $q$ is odd, the Chern number of each band, except the middle band, is -1. For $\alpha \ll 1$, the Berry curvature of these bands is increasingly uniform,  $\Omega_n ({\bf p}) \simeq \Omega_n$\cite{zakberry, cominotti, harper2014perturbative}. The average value, $|\Omega_n| = a^2 / (2 \pi \alpha)$, is estimated by noting that the Chern number: $\mathcal{C}_n = \frac{1}{2\pi }\Omega_nA_{\rm BZ} = -1$, where $A_{\rm BZ}= (2 \pi)^2 / q a^2$ is the area of the MBZ\cite{pricemodes}. Therefore for $\alpha = 1/q \ll 1$, the effective Hamiltonian (\ref{eq:key}) describes a particle in a uniform magnetic field on a torus in momentum space, with an additional overall energy shift.  

From the duality between real space and momentum space magnetism, we can translate known analytical results for (\ref{eq:magh}) to find the eigenspectrum and eigenstates of (\ref{eq:key}) (Supplemental Material). In a real space uniform magnetic field, the eigenstates are Landau levels\cite{Landau1981Quantum}. Restricting the particle to the surface of a torus, the infinitely degenerate Landau levels are superposed to satisfy the appropriate boundary conditions\cite{jain, AlHashimi}. 

Including the different Harper-Hofstadter bands, the resulting eigenspectrum of our model can be summarized as a collection of intertwined semi-infinite ladders,
\begin{eqnarray}
\mathcal{E}_{n,\beta} = E_n+\left[\beta + \frac{1}{2} \right] \kappa | \Omega_n |.  \label{eq:en}
\end{eqnarray} 
Each ladder starts at the energy $E_n$ of the band. Within each ladder, the states are classified by the Landau level quantum number $\beta=0,1,2,\ldots$ and their constant spacing is set by the analogue $\kappa|\Omega_n|$ of the cyclotron frequency $\omega_c=e|B|/M$. This is the well-known Landau level spectrum, unaffected by the toroidal topology. However, the topology does reduce the degeneracy of states from an infinite to a finite number, equal to the number of magnetic flux quanta inside the torus\cite{jain}. Counting this degeneracy may provide another experimental tool to directly measure the Chern number of nondegenerate bands. 

We also translate toroidal Landau levels from real space~\cite{AlHashimi} to the MBZ to find the expected analytical eigenstates (Supplemental Material). The Landau levels are strongly affected by global topology as, for example, their form depends on the Chern number, and the toroidal boundary conditions break translational symmetry in momentum space. 

{\it Comparison of numerical \& analytical eigenspectra.}-- Our analytical interpretation is confirmed by numerics, as shown in Fig. \ref{fig:energyspacing} over a suitable range of $\alpha$. Landau levels in the lowest band are spaced by $\kappa |\Omega_0| = \kappa a^2 / (2 \pi \alpha)$ (\ref{eq:en}). Multiplying by $\alpha$, this energy spacing is a constant (equal to $\simeq 0.003 J$ for $\kappa a^2 / J =0.02$). Numerically, this behavior is represented by the almost flat, equispaced states that are visible in Fig. \ref{fig:energyspacing} around $\alpha=0.1$. The level spacing was noted in Ref.~\onlinecite{kolovskypara} but its origin was not discussed. The eigenstates are nondegenerate as $|C_0|=1$. 

At higher energies in Fig. \ref{fig:energyspacing}, a second ladder of states cuts across the first. These can be identified as Landau levels in the second lowest Harper-Hofstadter band. (As the spacing is calculated relative to $\mathcal{E}_g$, only states from the lowest Harper-Hofstadter band are horizontal.) The strength of anticrossings between different states is controlled by band mixing from the external harmonic trap. For a sufficiently weak trap and a large band gap, the single-band approximation is valid and levels originating from different bands freely cross without significant coupling. This describes, for example, $\kappa a^2 / J =0.02$ at $\alpha  =1/11$, where the band gap is $(E_1 - E_{0}) /J \simeq 1$, and the effective Hamiltonian applies to each band separately. The breakdown of this behavior for a stronger trap is discussed in the Supplemental Material.

As $\alpha \rightarrow 0$, the Harper-Hofstadter bands become too close and band mixing is important. In this limit, the energies are those of a 2D simple harmonic oscillator on a tight-binding lattice\cite{kolovskypara}. As this is independent of $\alpha$, the numerical quantity plotted in Fig. \ref{fig:energyspacing} vanishes for all states. In the opposite limit, as $\alpha \gtrsim 0.2$, the energy spacing becomes distorted as we can no longer approximate $\Omega_n({\bf p})$ and $E_n({\bf p})$ as uniform: two assumptions which simplified the effective Hamiltonian. 

Note that although our analytics are restricted to $\alpha=1/q$, numerically the spectrum continuously depends on $\alpha$. The analytical explanation of this in the general $p>1$ case requires application of the magnetic model to (almost) degenerate bands with non-Abelian Berry connection, which will be the subject of a future publication.

\begin{figure}[!]
  \centering
  $
  \begin{array}{ccc}
(a)\resizebox{0.19\textwidth}{!}{\includegraphics*{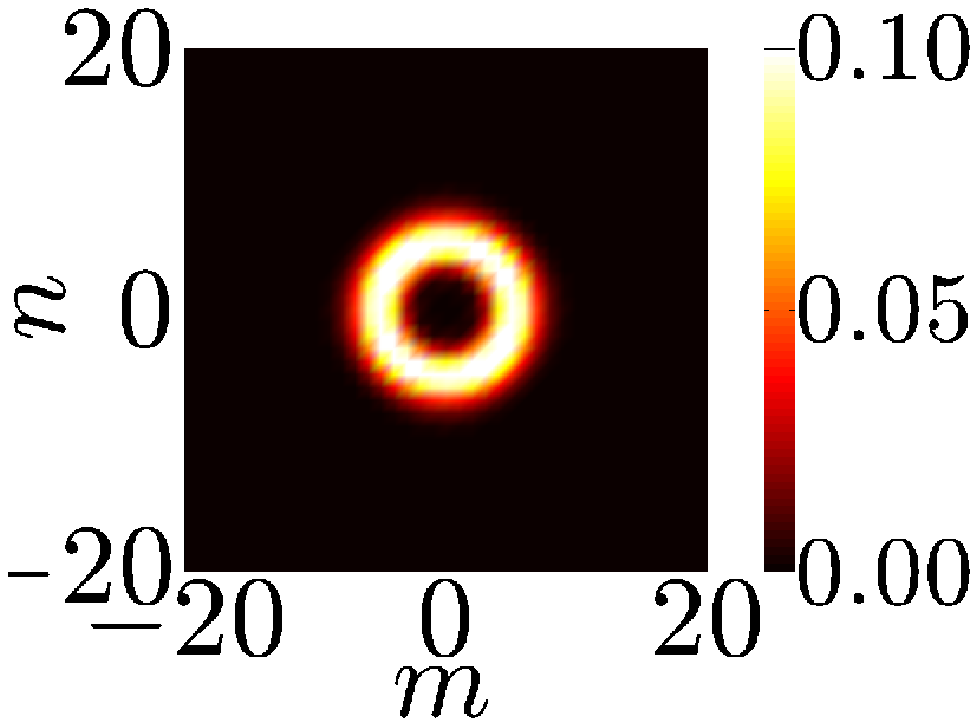}}  
& 
(b)\resizebox{0.138\textwidth}{!}{\includegraphics*{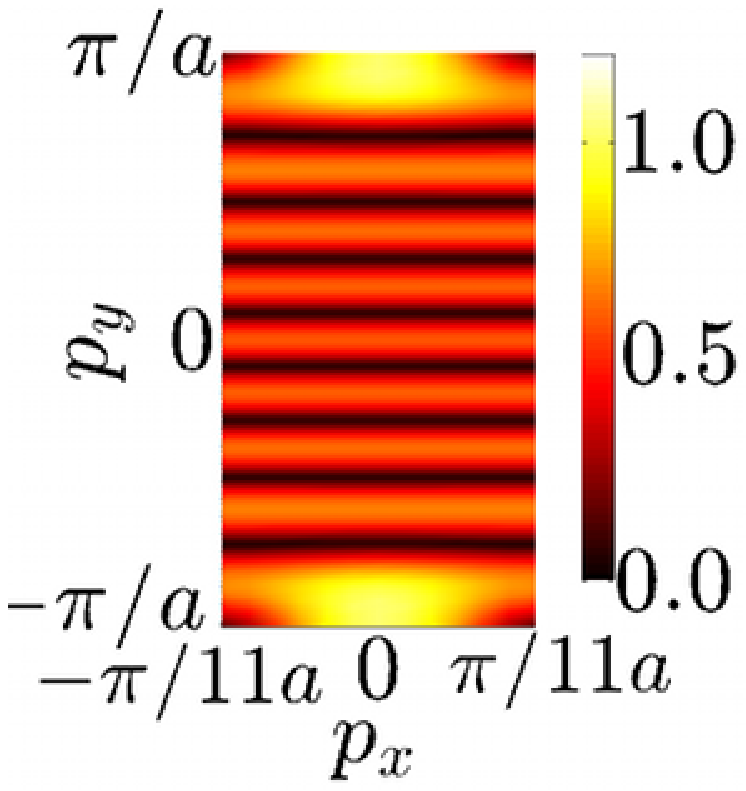} } 
\\
(c)\resizebox{0.19\textwidth}{!}{\includegraphics*{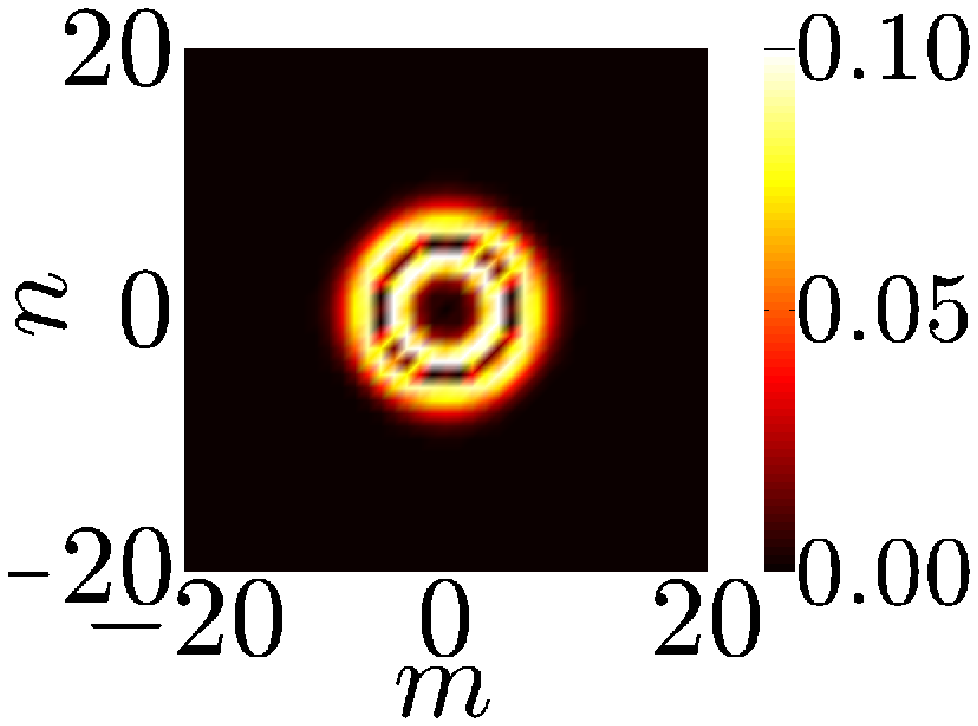} }
&
(d)\resizebox{0.138\textwidth}{!}{\includegraphics*{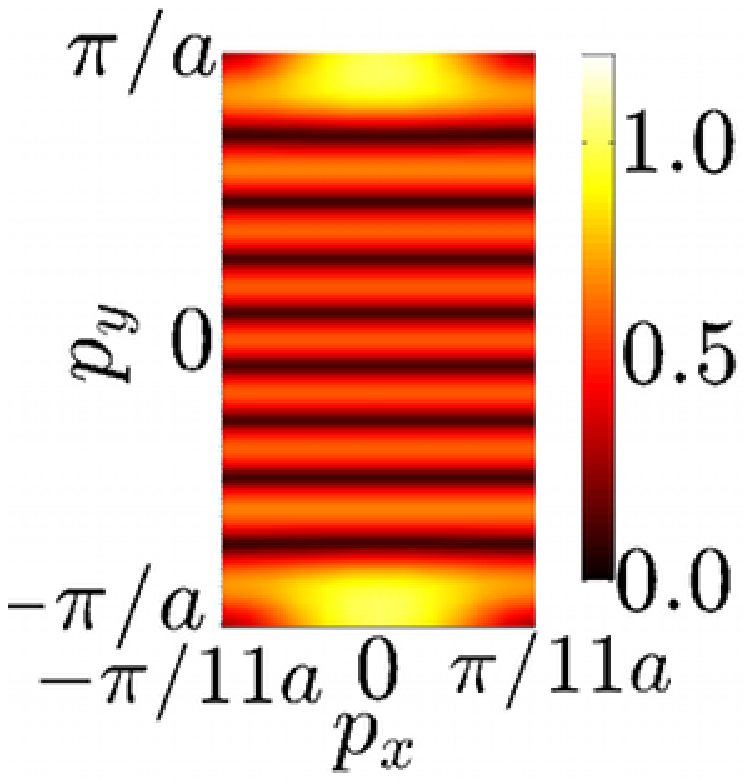} }
  \end{array} \newline
  \begin{array}{cc} 
&
(e)\resizebox{0.335\textwidth}{!}{\includegraphics*{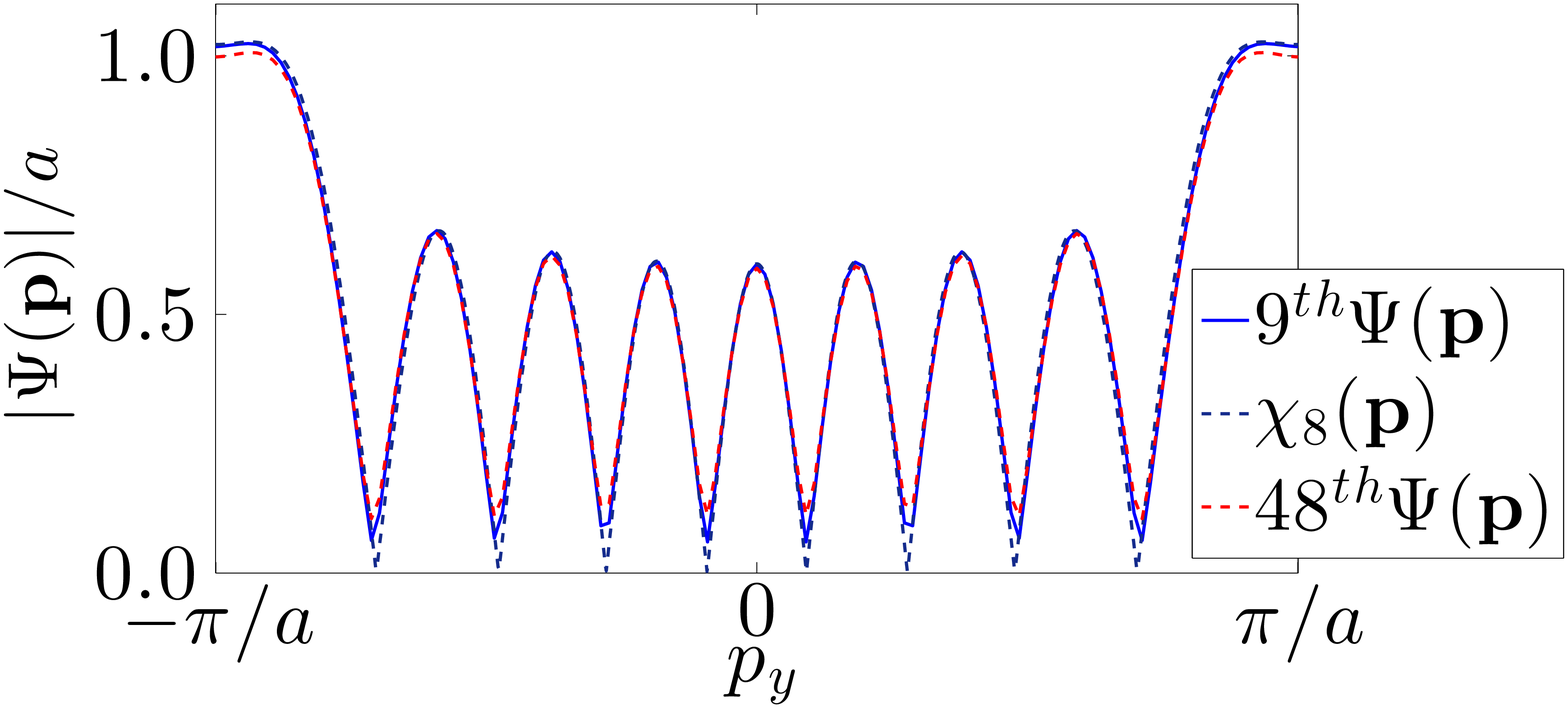}}  
\end{array} $
\caption{The top (middle) row is the 9th (48th) lowest eigenstate of Eq. (\ref{eq:hof}) for $\alpha = 1/11$, $\kappa a^2 /J =0.02$ and a lattice of $N=133\times 133$ sites. (a)\&(c) The numerical wave function, $|\psi({\bf r})|$, in units of $a^{-1}$. (b)\&(d) The numerical population over bands in the MBZ in units of $a$. (b)\&(d) are qualitatively indistinguishable from the analytical $\beta=8$ toroidal Landau level, $\chi_8 ({\bf p})$, in the lowest and second lowest Harper-Hofstadter band respectively. In both bands, $|\Omega| = a^2 / (2 \pi \alpha)$ and toroidal Landau levels have the same form (Supplemental Material). (e) Quantitative comparison along $p_x =0$ taken from (b),(d) \& the analytical $\beta=8$ level.}
 \label{fig:illustrate} 
\end{figure} 

{\it Comparison of numerical \& analytical eigenstates.}-- Figure \ref{fig:illustrate} shows the 9th \& 48th numerical states in real space (a)\&(c), and as a population over energy bands in the MBZ (b)\&(d). The latter are in excellent agreement with the analytical $\beta =8$ toroidal Landau level in the $n=0,1$ Harper-Hofstadter bands, as demonstrated quantitatively in Fig.~ \ref{fig:illustrate}(e). This agreement supports the single-band approximation, as our analytical wave function is able to capture all numerical features. 

The population in the MBZ is mostly determined by the Landau level quantum number $\beta$, with a number of nodes that appears to increase with $\beta$ as expected. The population is also nearly identical in the two bands (see panels (b)\&(d)). Conversely, the real space wave functions, shown in panels (a)\&(c), are markedly different, for example, with more nodes as the band index, $n$, increases. This difference is because the real space states depend on the $n$ via the Bloch wave functions $u_{n, \mathbf{p}} ({\bf r})$. 

{\it Experimental considerations.}--We observe that the form of the Landau levels is remarkably robust to parameter variation, making our proposal well suited to experimental investigation. According to our numerics, the basic features of the lowest energy toroidal Landau levels survive up to $\alpha=1/3$, provided that the harmonic trapping strength is larger than the bandwidth of the lowest band. Importantly, these results are also very insensitive to lattice size, due to strong localisation of the low energy eigenstates in real space (Fig. \ref{fig:illustrate}). This is because toroidal Landau levels vary over a large characteristic momentum scale, $l_{\Omega_n}= \sqrt{1/|\Omega_n |}$, which depends only on $\alpha$ and $\mathcal{C}_n$, and which is the analogue of the ``magnetic length", $l_{B}= \sqrt{1/ e|B|}$. For relevant parameters here (as well as typical parameters in more general systems), the wave function is delocalized in momentum, and hence localized in real space (Supplemental Material). 
 
To realize the proposed experiment, a harmonic trap can be straightforwardly added to an ultracold gas using additional laser beams and/or magnetic fields. The momentum space eigenstate structure can be probed directly in time-of-flight measurements of the momentum distribution when both the lattice and artificial magnetic field responsible for the complex hopping terms in (\ref{eq:hof}) are suddenly switched off. While the real space wave function (Fig. \ref{fig:illustrate} (a)\&(c)) is independent of the magnetic gauge choice, the momentum space wave function is not. However, in this experimental procedure the canonical momentum is measured directly as the physical momentum in the final time-of-flight expansion stage\cite{linelectric,aidelsburger, hiro}. In photonics, similarly, a harmonic potential can be created in the cavity arrays of Refs.~\onlinecite{hafezi2} \& \onlinecite{jacqmin} by letting the cavity size vary spatially, while the real (momentum) space wave function can be extracted from the near-field (far-field) emitted light~\cite{rmp2013,jacqmin}. 

{\it Summary.}-- We have introduced how future experiments may use external potentials and geometrical energy bands to design novel magnetic Hamiltonians in momentum space. As a first step, we have shown how a particle in a uniform magnetic field confined to a torus may be realized experimentally. The global toroidal topology has important consequences, for example, in the degeneracy of the eigenspectrum, in the spontaneous breaking of translational symmetry in momentum space and in the form of eigenstates. \\

\acknowledgments{We are grateful to N. R. Cooper for helpful comments and to P. Ghiggini for mathematical support. This work was partially funded by ERC through the QGBE grant and by the Autonomous Province of Trento, Call ``Grandi Progetti 2012," project ``On silicon chip quantum optics for quantum computing and secure communications - SiQuro".}

\end{document}

\section*{Supplementary Information}

\appendix 

\section{Derivation of The Effective Hamiltonian for an External Harmonic Trap} \label{app:harm}

Here we explicitly derive the effective Hamiltonian for an external harmonic trap, $W({\bf r}) = \frac{1}{2} \kappa {\bf r}^2$. We begin from the Schr\"{o}dinger equation, expanded in terms of the eigenstates of $\mathcal{H}_0$: 
\begin{eqnarray}
	i\frac{\partial}{\partial t}\psi_{n} ({\bf p}) &&
	=
	E_n (\mathbf{p}) \psi_{n}  ({\bf p})+	 \sum_{n} \sum_{\bf p} \frac{\kappa }{2 }  \langle \chi_{n^\prime,\mathbf{p}^\prime}| {\bf r}^2 |\chi_{n,\mathbf{p}}\rangle \psi_{n} ({\bf p}) \nonumber.
\end{eqnarray} 
In the last line, we insert the completeness relation of Bloch states, $1=  \sum_n   \sum_{\bf p} |\chi_{n,{\bf p}}\rangle \langle \chi_{n,{\bf p}}| $: 
\begin{eqnarray}
	i&\frac{\partial}{\partial t}&\psi_{n}(\mathbf{p})
	=
E_n (\mathbf{p}) \psi_{n}(\mathbf{p}) \nonumber \\
&&+	\frac{\kappa}{2 }  \sum_{n, n^{\prime \prime}}
	 \sum_{{\bf p}, {\bf p}^{\prime \prime}} \langle \chi_{n^\prime,\mathbf{p}^\prime}|  \mathbf{r} | \chi_{n^{\prime \prime},\mathbf{p}^{\prime \prime}}\rangle
\langle \chi_{n^{\prime \prime},\mathbf{p}^{\prime \prime}}| \mathbf{r} |\chi_{n,\mathbf{p}}\rangle \psi_{n} (\mathbf{p}) \nonumber
\end{eqnarray} 
Now that the Schr\"{o}dinger equation is in this form, we can apply an identity to replace terms of the type: $\langle \chi_{n^\prime, \mathbf{p}^\prime}|\mathbf{r}|\chi_{n,\mathbf{p}}\rangle$. This identity is a generalised version of a result originally due to Karplus and Luttinger\cite{karplus}. The derivation is as follows: 
\begin{align}
	\langle \chi_{n^\prime, \mathbf{p}^\prime}|&\mathbf{r}|\chi_{n,\mathbf{p}}\rangle
		=
	\int d^2 {\bf r}\  w^*_{n^\prime,\mathbf{p}^\prime}(\mathbf{r}) e^{-i\mathbf{p}^\prime\cdot \mathbf{r}} \mathbf{r}  e^{i\mathbf{p}\cdot \mathbf{r}}w_{n, \mathbf{p}}(\mathbf{r})
	\notag \\
	&=
	- i \int d^2 {\bf r} w^*_{n^\prime,\mathbf{p}^\prime}(\mathbf{r}) e^{-i\mathbf{p}^\prime\cdot \mathbf{r}} \left({\bm \nabla}_{\bf p} e^{i\mathbf{p}\cdot \mathbf{r}}\right)w_{n, \mathbf{p}}(\mathbf{r})	
	\notag \\
	&=
	-i\nabla_{\mathbf{p}} \left[ \int d^2 {\bf r}\ e^{-i(\mathbf{p}^\prime-\mathbf{p})\cdot \mathbf{r}}w^*_{n^\prime, \mathbf{p}^\prime}(\mathbf{r}) w_{n, \mathbf{p}}(\mathbf{r})\right] \nonumber \\
&+ i 
	\int d^2 {\bf r}\ e^{-i(\mathbf{p}^\prime-\mathbf{p})\cdot \mathbf{r}}w^*_{n^\prime, \mathbf{p}^\prime}(\mathbf{r}) \nabla_{\mathbf{p}} w_{n, \mathbf{p}}(\mathbf{r})   ,
\end{align}
where $w_{n,\mathbf{p}}(\mathbf{r})  = \langle {\bf r} | n {\bf p} \rangle$. If $\mathcal{H}_0$ is periodic, the momentum is the crystal momentum and $w_{n, \mathbf{p}}(\mathbf{r})   = u_{n, \mathbf{p}}(\mathbf{r})$. As the Bloch functions are periodic, the last integral will vanish\cite{karplus} unless ${\bf p} = {\bf p}^\prime$. Otherwise, for a free space $\mathcal{H}_0$, $w_{n, \mathbf{p}}(\mathbf{r}) =w_{n, \mathbf{p}}$ is independent of position. Then the last integral will also vanish unless ${\bf p} = {\bf p}^\prime$. Using the orthonormality of the eigenfunctions, and integrating the derivative of the delta function by parts, we obtain the identity:
\begin{align}
\langle \chi_{n^\prime, \mathbf{p}^\prime}|\mathbf{r}|\chi_{n,\mathbf{p}}\rangle
	&=
		\delta_{\mathbf{p},\mathbf{p}^\prime}
	\left(\delta_{n,n^\prime} i\nabla_{\mathbf{p}^\prime}	 +
	 i  \langle n^\prime{\bf p}^\prime| \frac{\partial}{\partial {\bf p}} | n{\bf p}\rangle\right) . \label{eq:abelian}
\end{align}
We can recognise the last term as the generalised non-Abelian Berry connection\cite{di}: $\boldsymbol{\mathcal{A}}_{n^\prime, n}(\mathbf{p}) = i  \langle n^\prime{\bf p}| \frac{\partial}{\partial {\bf p}} |n{\bf p}\rangle$. Substituting the identity into the Schr\"{o}dinger equation, we find:
\begin{align}
	i\frac{\partial}{\partial t}\psi_{n} ({\bf p})
	=&
E_n (\mathbf{p}) \psi_{n} ({\bf p}) 
+\frac{\kappa}{2}
	 \sum_{n^\prime, n^{\prime \prime}}[
	\left(\delta_{n^\prime,n^{\prime \prime}} i\nabla_{\mathbf{p}}
	+
	\boldsymbol{\mathcal{A}}_{n^\prime,n^{\prime \prime}}(\mathbf{p})
	\right) \nonumber \\& \cdot
	\left(\delta_{n^{\prime \prime},n} i \nabla_{\mathbf{p}}
	+
	\boldsymbol{\mathcal{A}}_{n^{\prime \prime},n}(\mathbf{p})
	\right)
	\psi_{n} ({\bf p}) ]. \label{eq:general}
\end{align}
If we assume that the external potential does not significantly mix energy bands, we can make a single band approximation. The Schr\"{o}dinger equation then reduces to: 
\begin{align}
	i\frac{\partial}{\partial t}\psi_{n} ({\bf p})
	=
	\left[
	E_n (\mathbf{p}) + \frac{\kappa}{2}
	\left(i \nabla_{\mathbf{p}}
	+
	\boldsymbol{\mathcal{A}}_{n}(\mathbf{p})
	\right)^2	\right]
	\psi_{n} ({\bf p}).
\end{align}
We note that the Schr\"{o}dinger equation could equally have been restricted to a sub-space of degenerate or nearly degenerate bands that are well-separated in energy from other bands. Then the sums in Eq. \ref{eq:general} only run over this sub-space of bands. The corresponding Berry curvature has a non-Abelian gauge structure\cite{wilczek, bliokh2005spin}. Physically, an energy band Hamiltonian containing the non-Abelian Berry connection, $\boldsymbol{\mathcal{A}}_{n^\prime, n}(\mathbf{p})$, may describe quantum spin Hall models, graphene or topological insulators\cite{hasankane, qizhang} in additional potentials.

\section{Toy Model of a Lattice with an External Harmonic Trap} \label{app:harm}

In this section, we briefly present the toy model of a lattice overlaid with an additional harmonic trap, as a simple example of a momentum space Hamiltonian. We assume the energy bands of the lattice have no geometrical Berry curvature, i.e. there is no momentum space magnetic field. We also assume that the harmonic trapping strength is much less than the band gap to justify the single-band approximation. Then the effective Hamiltonian is: 
\begin{eqnarray}
\mathcal{H} = 
	E_n (\mathbf{p}) + \frac{\kappa}{2}
	\left(i \nabla_{\mathbf{p}} \right)^2 ,
\end{eqnarray}
which corresponds to the Hamiltonian of a particle confined in an external potential, $E_n (\mathbf{p})$. If the harmonic trapping strength is much larger than the bandwidth of this energy band, then the external potential can be approximated as uniform $E_n (\mathbf{p}) \simeq E_n$, and the Hamiltonian is that of a free particle in momentum space. 

If instead the bandwidth is large compared with the harmonic trapping strength, we cannot neglect the variation of $E_n (\mathbf{p})$ in momentum space. Instead, let us assume that there is a single minimum in the bandstructure at ${\bm p}= {\bm 0}$, and apply the effective mass approximation. Then the Hamiltonian can be written as: 
\begin{eqnarray}
\mathcal{H} = 
	E_n ({\bm p}= {\bm 0})+ \frac{{\bm p}^2}{2 M^*} + \frac{\kappa}{2}
	\left(i \nabla_{\mathbf{p}} \right)^2 ,
\end{eqnarray}
where $M^* = \left(\frac{\partial^2 E_n }{ \partial p^2}\right)^{-1}$ is the effective mass of the energy band (assumed to be isotropic). This is the Hamiltonian of a particle in a 2D harmonic trap in momentum space. The solution of the equivalent real space Hamiltonian is well-known, and can be simply translated from real space to momentum space. The momentum space eigenstates are therefore 2D harmonic oscillator levels and the eigenspectrum is: 
\begin{eqnarray}
\mathcal{E}_\beta = E_n ({\bm p}= {\bm 0})+  ( \beta + 1)\omega'
\end{eqnarray}
where $\beta$ is the 2D quantum number and $\omega' =  \sqrt{\kappa / M^*}$ is the harmonic trapping energy in momentum space. This energy spectrum captures, as can be expected, the frequency, $\omega'$, of the dipole mode of a Bose-Einstein condensate in a 2D optical lattice in the presence of a harmonic trap\cite{kramer}.

\section{Numerical Calculations for the Harper-Hofstadter Model with a Harmonic Trap}

This section contains all details of how numerical results were obtained for the Harper-Hofstadter model with a harmonic trap. 

\subsection{General Numerical Procedure}\label{app:trans}

We begin from the full tight-binding Hamiltonian (Eq. 4 in the main text), written as a matrix in the localised lattice site basis $(\hat{a}^\dagger_{m,n} | 0 \rangle)$. (For a lattice of $N$ sites, this matrix has a dimension of $N \times N$.) The eigenstates of the Schr\"{o}dinger equation, $\mathcal{H} |\Psi\rangle = E |\Psi\rangle$, can be expanded as:
\begin{eqnarray}
|\Psi\rangle&=& \sum_{m,n} \psi({m,n}) a^\dagger_{m,n}|0\rangle, \label{eq:eigen}
\end{eqnarray}
in this basis, where $\psi({m,n})$ are the eigenstate coefficients at each lattice site. (The eigenstates are normalised to an area $Na^2$.) We then diagonalize the full Hamiltonian matrix to find the numerical eigenvalues and real space eigenstates. 

The only significant source of error in the numerical diagonalization is  from the restriction of the basis set to a finite lattice. To ensure this error is negligible, we have checked that the low-energy eigenspectrum converges for these system sizes to within the accuracy shown. This convergence is fast due to the localisation of low-energy real space eigenstates for the range of $\alpha$ that we consider (discussed both in the main text and below). 

\subsection{Extracting the Eigenstates in the Magnetic Brillouin Zone from Numerics}\label{app:trans}

We show how the numerical eigenstates within the magnetic Brillouin zone can be extracted from the real space eigenstates (Eq. \ref{eq:eigen}) obtained in diagonalizing the Hamiltonian. In the first step, we Fourier transform the real space wave function to give:
\begin{eqnarray}
|\Psi\rangle= \sum_{\tilde{p}_x, \tilde{p}_y} \psi(\tilde{p}_x,\tilde{p}_y) a^\dagger_{\tilde{p}_x, \tilde{p}_y}|0\rangle,  \label{eq:p2}
\end{eqnarray}
where $\tilde{{\bf p}} = (\tilde{p}_x, \tilde{p}_y)$ is the crystal momentum in the {\it full} Brillouin zone: $-\pi/a <  \tilde{{\bf p}} \leq \pi /a$.

We note that the magnetic Brillouin zone is defined from the Harper-Hofstadter Hamiltonian without a harmonic trap\cite{kohmoto, 1chang}. For clarity, we demonstrate this by taking the Fourier transform of the lattice creation and annihilation operators: 
\begin{eqnarray}
\hat{a}_{m, n} &=& \sum_{{\tilde{p}_x, \tilde{p}_y}} e^{ i   \tilde{p}_x m a + i \tilde{p}_y n a }  \hat{a}_{\tilde{p}_x, \tilde{p}_y}, \nonumber \\
\hat{a}^{\dagger}_{m, n} &=& \sum_{{\tilde{p}_x, \tilde{p}_y}} e^{ -i   \tilde{p}_x m a - i \tilde{p}_y n a}  \hat{a}^{\dagger}_{\tilde{p}_x, \tilde{p}_y},
\end{eqnarray}
and substituting them into $\mathcal{H}_0$. After simplifying the algebra, the Harper-Hofstadter Hamiltonian in the full Brillouin zone becomes\cite{bernevig2013}: 
\begin{eqnarray}
\mathcal{H}_0 &&= \sum_{{\tilde{p}_x, \tilde{p}_y}}[ -2 J  \cos(\tilde{p}_x a) \hat{a}^{\dagger}_{\tilde{p}_x, \tilde{p}_y} \hat{a}_{\tilde{p}_x, \tilde{p}_y} 
\nonumber \\ 
&&- J ( e^{ - i \tilde{p}_y a} \hat{a}^{\dagger}_{\tilde{p}_x + 2 \pi \alpha /a , \tilde{p}_y} \hat{a}_{\tilde{p}_x, \tilde{p}_y} +  e^{  i \tilde{p}_y } \hat{a}^{\dagger}_{\tilde{p}_x- 2 \pi \alpha /a, \tilde{p}_y} \hat{a}_{\tilde{p}_x, \tilde{p}_y})
] . \label{eq:Hp}  \nonumber
\end{eqnarray}
This is not diagonal in the full Brillouin zone, as $\tilde{p}_x$ is mixed with  $\tilde{p}_x + 2 \pi \alpha /a$ and $\tilde{p}_x- 2 \pi \alpha /a$. However, the Hamiltonian can be made diagonal, by defining a new variable: $p_x = \tilde{p}_x + j 2 \pi \alpha /a$, where $j$ is an integer. For $\alpha = 1/q$, ${\bf G} = 2 \pi \alpha /a \hat{p}_x$ is a magnetic reciprocal lattice vector for the unit cell discussed in the text, and ${\bf p}$ is the magnetic crystal momentum. Note that the choice of the new variable, ${\bf p}$, followed naturally from the magnetic gauge of $\mathcal{H}_0 $. We have deliberately picked the magnetic unit cell for which: ${\bf p} = \tilde{{\bf p}} + j {\bf G}$. (For other choices of magnetic unit cell, this relation would not generally take this simple form.) We can then write: 
\begin{eqnarray}
| \chi_{n,{\bf p}} \rangle =  \sum_{\tilde{{\bf p }}} U_{n, {\bf p}} (\tilde{ {\bf p}}) \hat{a}^\dagger_{{\tilde{{\bf p}}}}| 0 \rangle ,\label{eq:transform}
\end{eqnarray}
where $U_{n, {\bf p}} ( \tilde{{\bf p}})$ is a unitary matrix that transforms eigenstates between the full and magnetic Brillouin zone. This matrix only has non-zero values for $\tilde{{\bf p}} = {\bf p} - j {\bf G}$. Taking the inverse of Eq. \ref{eq:transform} and applying it to Eq. \ref{eq:p2}, we have:
\begin{eqnarray}
|\Psi\rangle&=& \sum_{n, {\bf p}, \tilde{{\bf p}}}  U^*_{n, {\bf p}} ( \tilde{\bf p}) \psi(\tilde{\bf p})| \chi_{n,{\bf p}} \rangle \nonumber \\ 
& =& \sum_{n, {\bf p}} \psi_n({\bf p})  | \chi_{n,{\bf p}} \rangle,  
\end{eqnarray} 
where $\psi_n({\bf p}) $ are the wave function coefficients in the magnetic Brillouin zone. Then it follows that: 
\begin{eqnarray}
\sum_n |\psi_n({\bf p})|^2 = \sum_{j} | \psi (\tilde{\bf p} = {\bf p} -  j {\bf G})|^2 . \label{eq:relation}
\end{eqnarray} 
From the numerical real space wave function, we can therefore extract the population over all bands as a function of momentum in the magnetic Brillouin zone. 

\section{Analytical Interpretation of the Harper-Hofstadter Model with a Harmonic Trap}

In the regime, $\alpha = 1/q \ll 1$, our interpretation is the clearest thanks to the two simplifications described in the main text. These are that the energy band dispersion only contributes an overall energy shift, $E_n ({\bf p}) \simeq E_n$, and that the Berry curvature is approximately uniform $\Omega_n ({\bf p}) \simeq \Omega_n$. Then the effective Hamiltonian of the lowest energy bands can be written as:
\begin{eqnarray}
	\tilde{\mathcal{H} }	&=& E_n+ 
	  \frac{\kappa
	\left(i {\bm \nabla}_{\mathbf{p}}
	+
	\boldsymbol{\mathcal{A}}_{n}(\mathbf{p})
	\right)^2}{2}, \label{eq:1}
\end{eqnarray} 
where ${\bm \nabla}_{\bm p} \times \boldsymbol{\mathcal{A}}_{n}(\mathbf{p}) \cdot \hat{{\bm z}} = \Omega_n$ and where $|\Omega_n| = a^2 / (2 \pi \alpha)$. In the magnetic analogy, this is equivalent to the real space Hamiltonian of a charged particle on a torus:  
\begin{eqnarray}
	{\mathcal{H}}	&=& 
	  \frac{
	\left(- i {\bm \nabla}_{\mathbf{r}}
	- e
	\boldsymbol{{A}}(\mathbf{r})
	\right)^2}{2M} + e\Phi \label{eq:2}
\end{eqnarray} 
where ${\bm \nabla}_{\bm r} \times\boldsymbol{{A}}(\mathbf{r})\cdot \hat{{\bm z}} = B$, and $e \Phi$ is an overall scalar energy shift. The properties of this Hamiltonian are known, as magnetism in nontrivial topological spaces has previously been of mathematical interest (e.g. \cite{AlHashimi}). As shown below, we translate these known results from real space to momentum space magnetism, to give the analytical predictions used in the main text. 

\subsection{Eigenspectrum}

As toroidal Landau levels can be viewed as superpositions of the usual infinitely degenerate Landau levels, the eigenspectrum of the real space magnetic Hamiltonian (\ref{eq:2}) is unaffected by the torus and is given by\cite{AlHashimi}: 
\begin{eqnarray}
\mathcal{E}_{\beta} =e\Phi  +\left[\beta + \frac{1}{2} \right] \omega_c,  \label{eq:en1}
\end{eqnarray} 
where $\beta=0, 1, 2...$ is the Landau level quantum number and $\omega_c = e |B| / M$ is the cyclotron frequency. From the correspondence between Eq. \ref{eq:1} and \ref{eq:2}, we identify the following substitutions between real space and momentum space magnetism: 
\begin{eqnarray}
e\Phi& \rightarrow &E_n \nonumber \\
e |B| &\rightarrow& |\Omega_n| \nonumber \\
 M &\rightarrow& \kappa^{-1} . \label{eq:subs}
 \end{eqnarray}
From this analogy, Eq. \ref{eq:en1} can be translated as:
\begin{eqnarray}
\mathcal{E}_{n, \beta} =E_n  +\left[\beta + \frac{1}{2} \right] \kappa |\Omega_n|,  \label{eq:en}
\end{eqnarray} 
as stated in the main text. 

\subsection{Eigenstates}

The eigenstates of the real space magnetic Hamiltonian (\ref{eq:2}) in the Landau gauge on a torus are\cite{AlHashimi}: 
\begin{eqnarray}
 \chi_\beta ( {\bf r}) &=& \mathcal{N}_\beta^{l_{B}}  \sum_{j = - \infty}^{\infty} e^{\frac{2 \pi  i  y}{L_y}( n_\phi j  + l_y + \frac{\theta_y}{2\pi})- i \theta_x j } 
 \nonumber \\
 &&
 \times
 e^{ - ( x  +  j  L_x  +\frac{l_y L_x}{n_\phi}  + \frac{\theta_y L_x}{2\pi n_\phi}     ) ^2 / 2 l_{B}^2}  \nonumber \\
&&\times  H_\beta \left(   \frac{ x + j  L_x + \frac{l_y L_x}{n_\phi}  + \frac{\theta_y L_x}{2\pi n_\phi}   }{l_B}  \right)   \label{eq:torusreal}
\end{eqnarray} 
where $H_\beta$ are Hermite polynomials; $n_\phi$ is the integer number of magnetic flux inside the torus; $l_B = \sqrt{1 / e |B|}$ is the characteristic ``magnetic length"; $L_x$ ($L_y$) is the length of the torus along the $x$($y$) directions; and $l_y \in {0, 1, ... , |n_\phi|-1}$. The variables $\theta_x$ and $\theta_y$ are a consequence of the toroidal boundary conditions, and break the translational symmetry of the wave function on the torus\cite{AlHashimi}. The normalisation constant, $\mathcal{N}_\beta^{l_{B}}$, is given by: 
\begin{eqnarray}
\mathcal{N}_\beta^{l_{B}} &=& \left( \frac{\sqrt{2 n_\phi L_x / L_y}} {(2^\beta \beta! \times 2 \pi l_{B}^2 n_\phi)} \right)^{1/2}. 
\end{eqnarray}

We translate these analytical results from real space to momentum space by identifying analogous quantities. In addition to the substitutions listed above (\ref{eq:subs}), we have:
\begin{eqnarray}
n_\phi & \rightarrow & \mathcal{C}_n \nonumber \\
L_x &\rightarrow& \frac{2 \pi}{ q a}\nonumber \\
L_y &\rightarrow& \frac{2 \pi}{  a}\nonumber \\
l_B &\rightarrow&l_{\Omega_n} \simeq \sqrt{ \frac{2 \pi }{ |\mathcal{C}_n| a^2 q}} ,
 \end{eqnarray}
where we have used that $L_x (L_y)$ correspond to the lengths of the MBZ in the $p_x (p_y)$ direction. This provides a ``dictionary" with which to translate the toroidal Landau level wave function into momentum space. It can also be shown that when the real space origin of the harmonic trap coincides with the position of a lattice site, the variables $\theta_y=\theta_x =0$ \footnote{T. Ozawa, H. M. Price and I. Carusotto, in preparation.}, breaking the translational symmetry in the MBZ. We will consider the case $\theta_y=\theta_x =0$ hereafter for simplicity. For the lowest energy bands with $\mathcal{C}_n=-1$, the analytical eigenstate is: 
\begin{eqnarray}
 \chi_\beta ( {\bf p}) &=& \mathcal{N}_\beta^{l_{\Omega_n}}  \sum_{j = - \infty}^{\infty} e^{- i  p_y j a} e^{ - ( p_x  + j a  l_{\Omega_n}^2   ) ^2 / 2 l_{\Omega_n}^2}  \nonumber \\
&&\times  H_\beta (   p_x / l_{\Omega_n} +  j a   l_{\Omega_n}) \label{eq:torusag}
\end{eqnarray} 
as $l_y=0$ and the normalisation constant is:
\begin{eqnarray}
\mathcal{N}_\beta^{l_{\Omega_n}} &=& \left( \frac{\sqrt{2/q}} {(2^\beta \beta! \times 2 \pi l_{\Omega_n}^2)} \right)^{1/2}. 
\end{eqnarray}
We note that, like ${\bf A} ({\bf r})$, the momentum space Berry connection is gauge-dependent, while the Berry curvature is not. In writing down this analytical wave function, we have chosen the Landau gauge for the Berry connection, $\boldsymbol{{\mathcal A }}_n ({\bf p}) = \Omega_n p_x \hat{{\bf p}}_y$, however, we only compare quantities that are independent of this gauge choice with numerics. 

\section{Further Comparisons Between Numerical Results and Analytical Interpretation}

\subsection{Eigenspectrum}

\begin{figure}[!]
  \centering
    $
  \begin{array}{c}
(a)\resizebox{0.47\textwidth}{!}{\includegraphics*{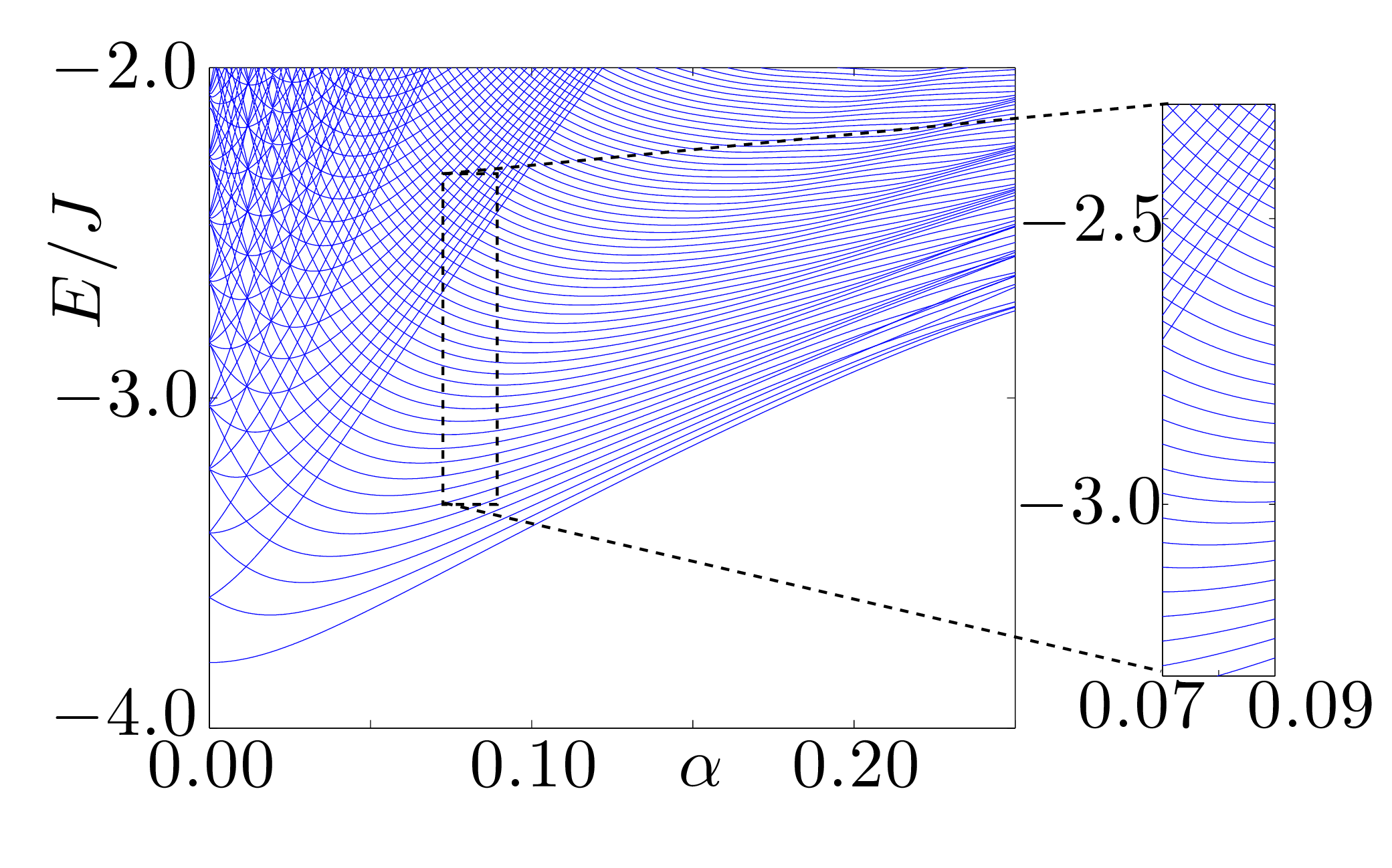}}    \\
(b) \resizebox{0.47\textwidth}{!}{\includegraphics*{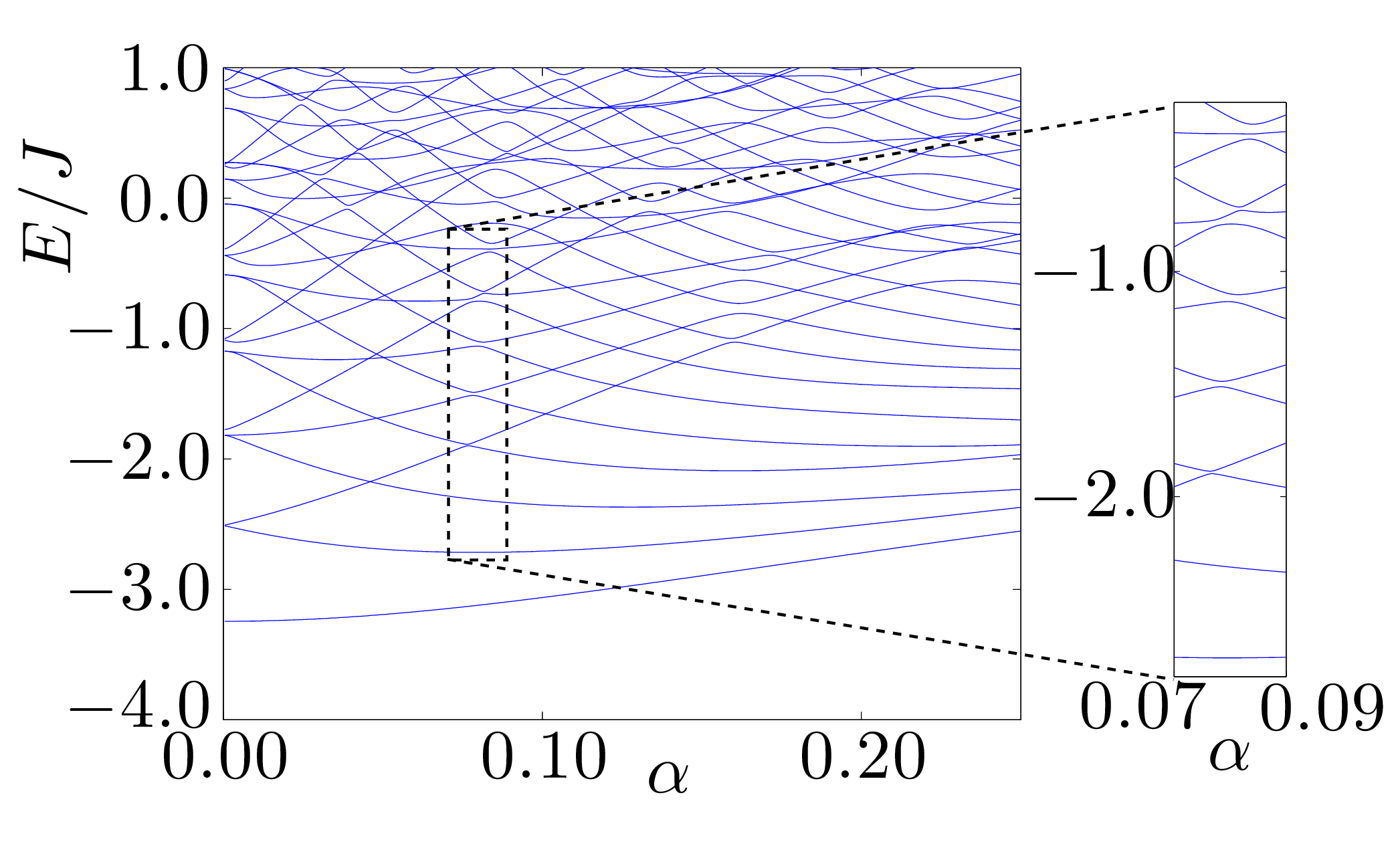}}   
\end{array}$
\caption{The low energy spectrum from numerically diagonalising Eq. 4 in the main text, for a lattice with $N= 41 \times 41$ sites, with (a) a weak trap: $\kappa a^2 / J = 0.02$  (b) a strong trap: $\kappa a^2 / J = 0.3$. The inset is an enlargement of the spectrum, emphasising the spacing of toroidal Landau levels and the degree of mixing between energy bands.} \label{fig:energyspacing} 
\end{figure}

In Figure E\ref{fig:energyspacing}, we show the unprocessed low energy numerical eigenspectrum of Eq. 4 in the main text for two values of the harmonic trapping strength. In the case of a weak trap, as shown in Fig. E\ref{fig:energyspacing}(a), the numerics are in good agreement with analytics over a wide-range of $\alpha$. This behaviour was discussed in the main text, and here we emphasise again some of the key points. 

Firstly, it is possible to estimate from Fig. E\ref{fig:energyspacing}(a) the range of validity of our interpretation. For the lowest energy states, for example, the Landau levels can be identified within the approximate range $0.02 \lesssim \alpha \lesssim 0.16$. As $\alpha \rightarrow 0$, the single band approximation breaks down and one recovers the energy spacing of approximately $\sqrt{2 \kappa a^2 / J}$ for a 2D simple harmonic oscillator on a tight-binding lattice\cite{kolovskypara}. Alternatively, as $\alpha \gtrsim 0.16$, our analytical assumption that the Berry curvature and the energy dispersion are both uniform breaks down, and the equal level spacing within the ladders of states is destroyed reflecting behaviour beyond the simple magnetic model (\ref{eq:2}). 

Secondly, when the trapping strength is sufficiently weak, there is no significant coupling between toroidal Landau levels in the different Harper-Hofstadter bands. This can be seen at higher energies in the inset of Fig. E\ref{fig:energyspacing}(a), where a second ladder of states cuts diagonally across the first. These states can be identified as toroidal Landau levels in the second lowest Harper-Hofstadter band; the similar spacing is a consequence of $\mathcal{C}_0=\mathcal{C}_1$, while the overall upwards shift is due to the energy difference $(E_1 - E_{0})$. The ladders in different bands are only weakly coupled as the harmonic trapping strength is very small compared with the band gap.

Increasing the harmonic trapping strength increases the energy level spacing, as shown in Fig. E\ref{fig:energyspacing}(b). Most importantly, the single-band approximation begins to break down over a wider range of $\alpha$. For example, for $\alpha=1/11$ and $\kappa a^2 / J = 0.3$, the harmonic trap is no longer small compared with the band gap and our analytical interpretation is less appropriate. As highlighted in the inset of Figure E\ref{fig:energyspacing}(b), increasing the trapping strength for a given value of $\alpha$, increases the coupling between different bands. This can be seen in the robust anti-crossings of states, that strongly perturb the energy level spacing.

\subsection{Determination of the Landau Level Indices for Numerical Eigenstates}

As shown in the main text, there is good agreement with the $\beta=8$ toroidal Landau level for both the 9th and 48th lowest numerical eigenstates when $\alpha=1/11$ and $\kappa a^2 / J =0.02$. Here we present explicitly how the relevant Landau level number, $\beta$, and band index, $n$, can be determined for the $k$-th numerical eigenstate. 

At low energies, $ \mathcal{E}\lesssim E_1$, all states belong to the ladder of toroidal Landau levels in the lowest band (Eq. \ref{eq:en} with $n=0$). In this regime, the $k$-th numerical eigenstate corresponds to a Landau level with:
\begin{eqnarray}
\beta=k-1,  \qquad n=0 \nonumber
\end{eqnarray}
as $\beta =0, 1,...$. This relation describes, for example, the $k=9$ numerical eigenstate which has $\beta =8$ and $n=0$, when $\alpha=1/11$ and $\kappa a^2 / J =0.02$. 

At higher energies, $E_2 \gtrsim\mathcal{E}\gtrsim E_1$, there is a ladder of Landau levels in the second lowest Harper-Hofstadter band in addition to the first ladder. As the two lowest bands both have $\mathcal{C}=-1$, the average Berry curvature is equal, $ |\Omega_1| \simeq  |\Omega_0|$. Consequently, Landau levels for the $n=0$ and $n=1$ Harper-Hofstadter bands have the same form for a given value of $\beta$ (\ref{eq:torusag}). The energy spacing of states within each ladder is also equal as $ \kappa |\Omega_1| \simeq  \kappa |\Omega_0|$. Then, as we increase the energy above the onset energy of the second ladder (\ref{eq:en}), states with $k$ and $k+1$ belong alternately to the ladders with $n=0$ and $n=1$. 

In this energy regime, we must identify where each numerical state lies within the appropriate ladder of Landau levels. By counting states, the indices are given by: 
\begin{eqnarray}
&& \beta = k_0 + \frac{k-k_0}{2}-1, \qquad n=0\qquad \mbox{if mod}(k-k_0, 2)=0 \nonumber \\
&& \beta = \frac{k-k_0-1}{2} , \qquad n=1\qquad \mbox{ if mod}(k-k_0, 2)=1\nonumber
\end{eqnarray}  
for $k> k_0$, where $k_0$ is the integer number of eigenstates lying below the onset energy of the second ladder. The number of states, $k_0$, can be estimated from the energy band gap, $E_1-E_0$, divided by the spacing of the energy levels. For $\alpha=1/11$ and $\kappa a^2 / J =0.02$, we find numerically that $k_0 = 31$, i.e. the first 31 states are in the $n=0$ Harper-Hofstadter band, while the 32nd state corresponds to the $\beta=0$, $n=1$ toroidal Landau level. From the above relations, the $k=48$ numerical eigenstate then has $\beta=8$ and $n=1$ as verified in the main text.

Note that, analytically, we would estimate $k_0 =29$ from Eq. \ref{eq:en} for these parameters. The discrepancy with numerics is because the energy level spacing is proportional to $\kappa a^2 / J$, and so is very small. Consequently, $k_0$ is large and sensitive to any corrections beyond our approximations; the discrepancy vanishes as we increase the harmonic trapping strength. We emphasise that this only affects how we identify the relevant indices for numerical states. For any given state, corrections to the Landau levels are small for this value of the trapping strength, as shown in the good agreement between numerical and analytical results in the main text and below. 

\begin{figure*}[!]
\centering
$
\begin{array} {ccc}
(a)\resizebox{0.21\textwidth}{!}{\includegraphics*{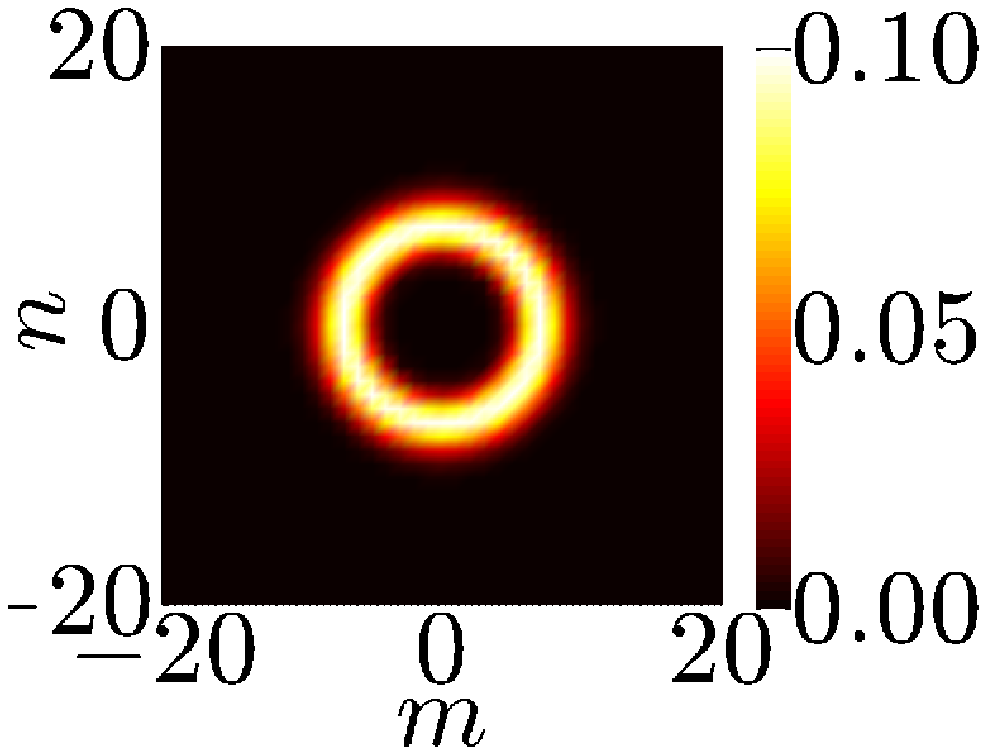}}  
& 
(b)\resizebox{0.21\textwidth}{!}{\includegraphics*{Real48_new2}}  
&
(c)\resizebox{0.21\textwidth}{!}{\includegraphics*{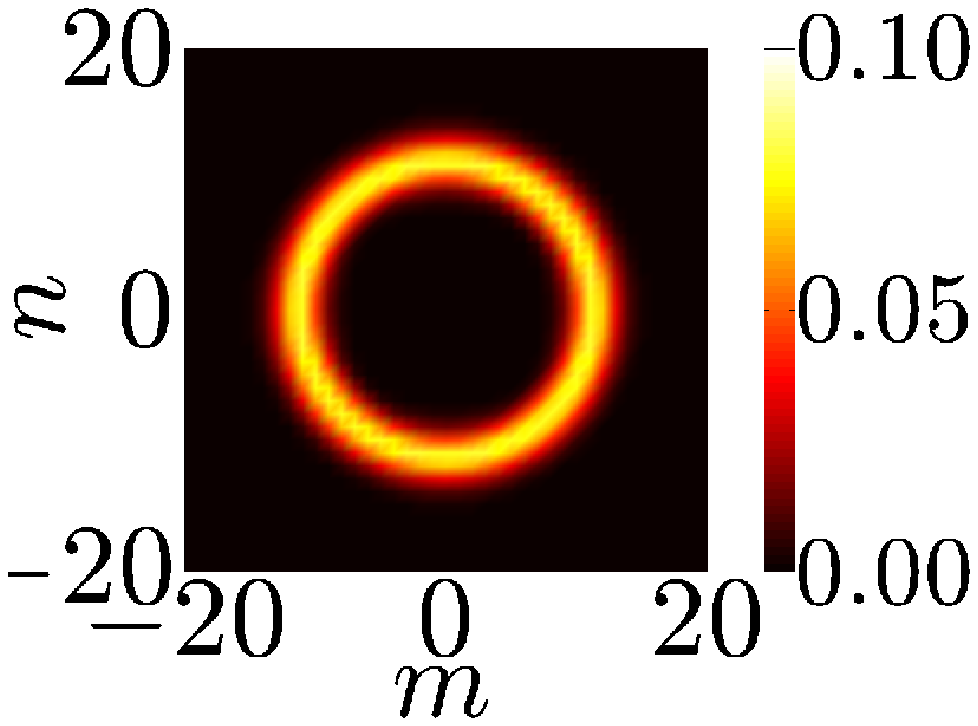}}  
\\
(d)\resizebox{0.15\textwidth}{!}{\includegraphics*{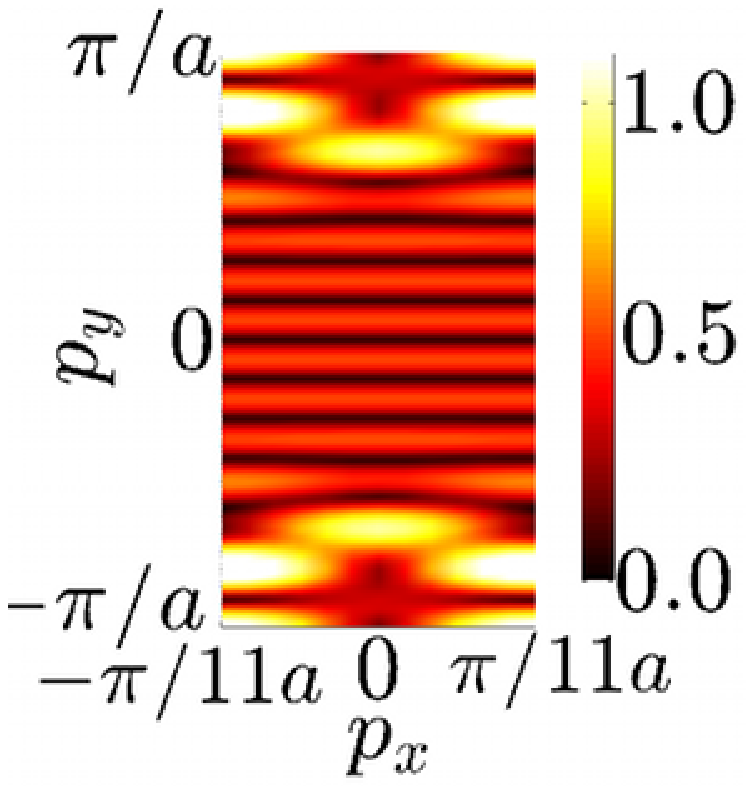} }
& 
(e)\resizebox{0.15\textwidth}{!}{\includegraphics*{Mag48th_3} }
&
(f)\resizebox{0.15\textwidth}{!}{\includegraphics*{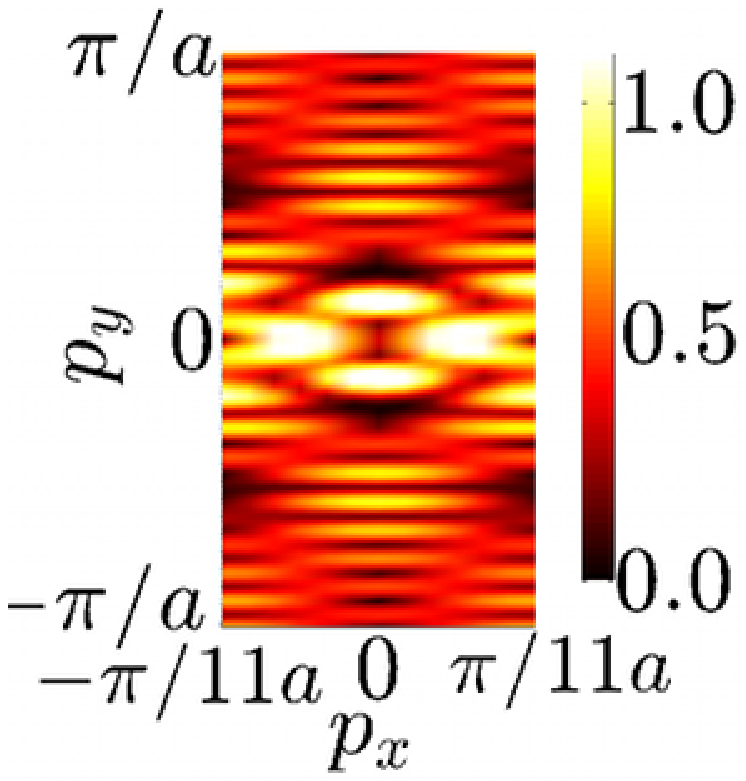} }
\\
(g)\resizebox{0.16\textwidth}{!}{\includegraphics*{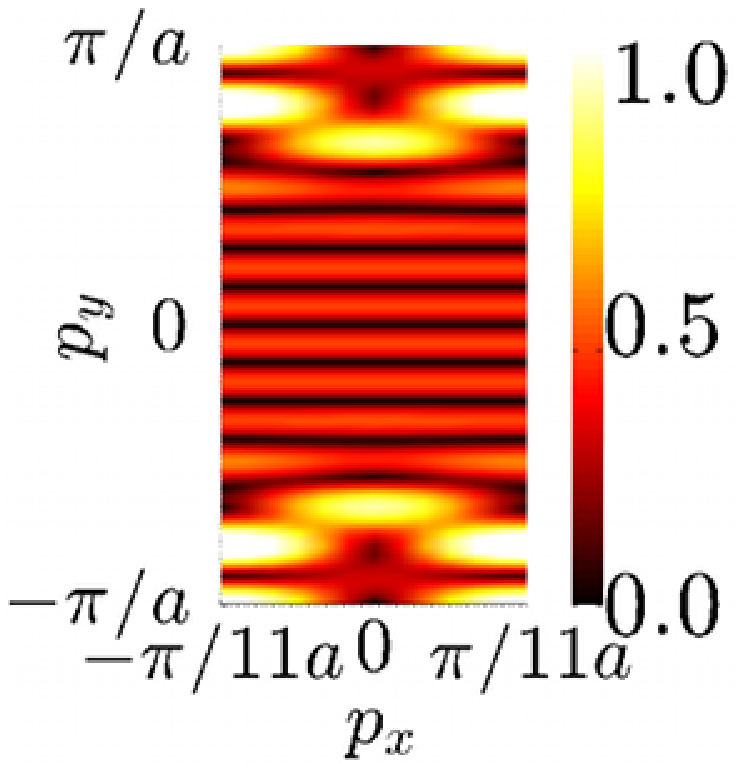} }
&
(h)\resizebox{0.16\textwidth}{!}{\includegraphics*{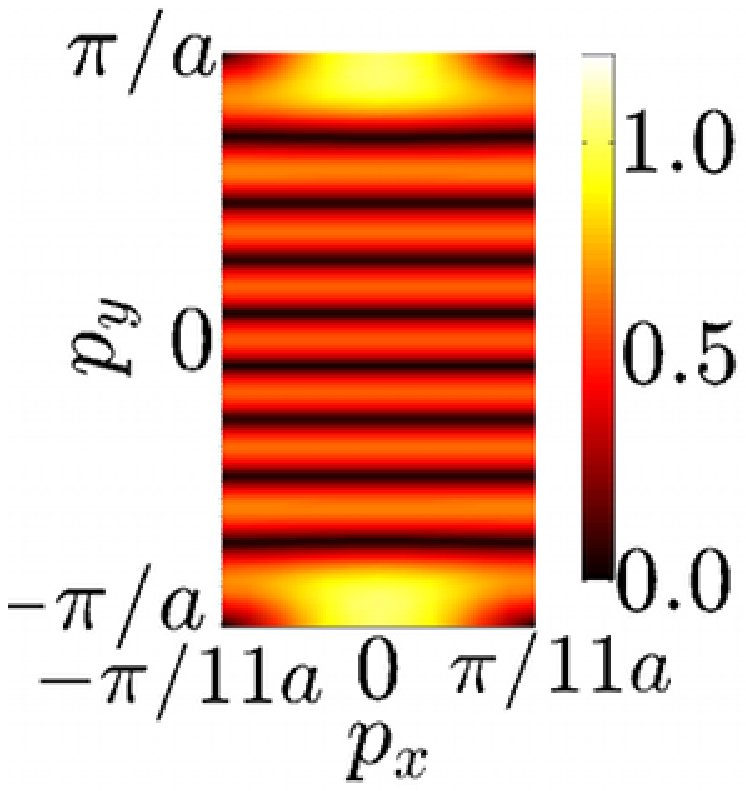} }
& 
(i)\resizebox{0.16\textwidth}{!}{\includegraphics*{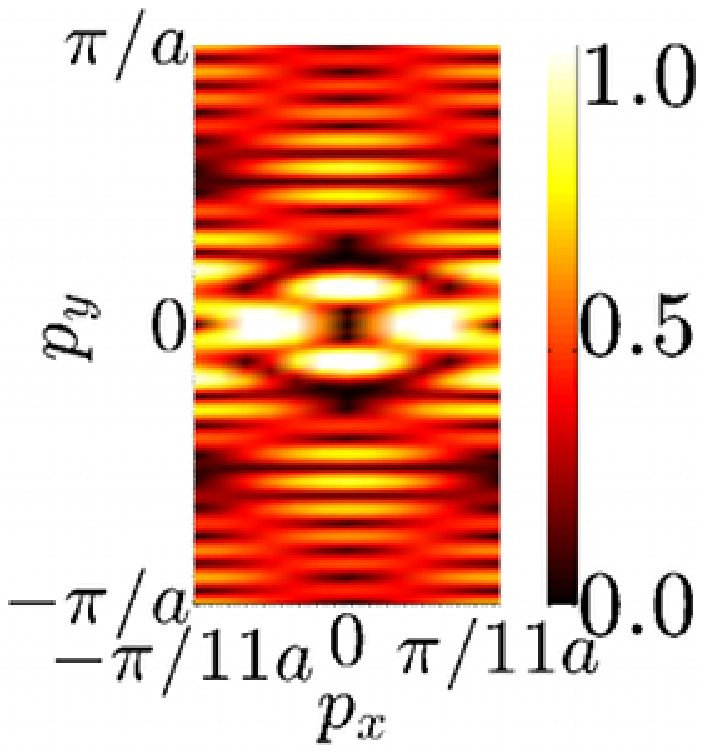} }

\end{array}$
\caption{(a),(b)\&(c) The wave function, $|\psi({\bf r})|$, in units of $a^{-1}$. From left to right, these are the 16th, 48th and 49th lowest eigenstate of the Harper-Hofstadter Hamiltonian with a harmonic trap for $\alpha = 1/11$, $\kappa a^2 /J =0.02$ and $N=(133)^2$. (The 48th eigenstate was also shown in the main text, and is included here to facilitate further comparisons.) (d),(e)\&(f) The population over all energy bands in the magnetic Brillouin zone, in units of $a$. The middle column can be identified as a Landau level in the second lowest Harper-Hofstadter band, while the other two columns correspond to Landau levels in the lowest Harper-Hofstadter band. (g),(h)\&(i) From left to right, the analytical $\beta=15$, $\beta=8$ \& $\beta=39$ toroidal Landau level (Eq. \ref{eq:torusag}) in units of $a$. These are in good agreement with numerical results (d),(e)\&(f). } \label{fig:landau} 
\end{figure*}

Finally, as the energy is increased still further, $\mathcal{E}\gtrsim E_2$, states from higher Harper-Hofstadter bands will also become important and the above relations require a (straightforward) generalization to include them. However, all example eigenstates discussed in this paper are below this energy scale. 

\subsection{Properties of Eigenstates}

Further examples of toroidal Landau levels are shown in Figure E\ref{fig:landau}, to demonstrate the exotic patterning of nodes in these eigenstates. Displayed here are the 16th, 48th and 49th numerical eigenstates. These correspond to toroidal Landau levels for $\alpha=1/11$ and $\kappa a^2 / J =0.02$ with $\beta=15$ \& $n=0$, $\beta=8$ \& $n=1$, and $\beta=39$ \& $n=0$ respectively. (The indices, $\beta$ and $n$, are determined as described above.) The numerical results for the MBZ are again in excellent agreement with the analytical toroidal Landau levels (\ref{eq:torusag}). 

The band index determines the number of nodes in the real space wave function through the Bloch function, $u_{n, {\bf p}}({\bf r})$, as can be seen by comparing panels (a),(b)\&(c) in Fig. E\ref{fig:landau}. In momentum space, the band index enters the form of the toroidal Landau level through the Chern number and average Berry curvature. However, in our model, these properties are the same in both the $n=0$ and $n=1$ bands, and so the analytical eigenstates are identical in the two bands (see also Fig. 2 in the main text). We now focus further on two specific aspects of the eigenstates; the consequence of toroidal topology for the Landau levels and the localisation/de-localisation of eigenstates in real and momentum space. 

\subsubsection{Consequences for Landau Levels of the Toroidal Topology of the MBZ }

To understand the consequences of toroidal topology for Landau levels, it can be useful to compare (\ref{eq:torusag}) with the standard form of Landau levels {\it not on a torus}:
\begin{eqnarray}
\chi_\beta ( {\bf p}) &\propto& e^{ i  p_y y} e^{ - ( p_x  -  y   l_{\Omega_n}^2   ) ^2 / 2 l_{\Omega_n}^2}   \nonumber \\
&&\times H_\beta (   p_x / l_{\Omega_n} -  y   l_{\Omega_n}) , \label{eq:nonto}
\end{eqnarray}
where ${\bf p}$ varies continuously over all momenta and is not constrained to the Brillouin zone. (The Berry connection is again chosen in the Landau gauge). Here we have translated again from real space magnetism to momentum space magnetism, but we note this discussion is meant illustratively and not with a particular underlying model in mind. 

The wave functions of these standard Landau levels are characterised by a plane wave along $p_y$, and $\beta$ nodes along $p_x$. In the same gauge, the toroidal Landau levels have a markedly different structure (Fig. E\ref{fig:landau} c \& d). The toroidal Landau levels vary along both momenta directions, with the majority of nodes along $p_y$. This is because the boundary conditions of the torus spontaneously break the translational symmetry in the MBZ along both directions [12].  We also note that while the number of nodes in toroidal Landau levels increases with $\beta$ as for usual Landau levels, it is more difficult to count these nodes definitively as the distribution of nodes is non-uniform along $p_x$ due to this symmetry breaking.

The standard Landau levels also only depend upon the magnetic field through the characteristic scale $ l_{\Omega_n}$, whereas on a torus, the wave function is also sensitive to $\mathcal{C}_n$, the {\it topological invariant} of the torus. Finally, whereas standard levels are infinitely degenerate, the toroidal Landau levels have a finite degeneracy equal to the Chern number, and so are also characterised by an additional quantum number $l_y$ which indexes the degenerate states.

\hspace{0.1in}
\subsubsection{ Localisation and Delocalisation of Eigenstates}

As introduced in the main text, the strong localisation of eigenstates in real space, for $\alpha = 1/q \ll1$, corresponds to the large characteristic momentum scale of Landau levels in momentum space: 
\begin{eqnarray}
l_{\Omega_n} = \sqrt{\frac{1 }{ |\Omega_n|}}\simeq \sqrt{\frac{2 \pi  }{ a^2 |\mathcal{C}_n| q}}
\end{eqnarray}
Like the analogous magnetic length in real space, this decreases as the magnetic field strength increases, reflecting the increased localisation of the wave function. However, for typical values of $\alpha$ for which our interpretation is valid, this is always still a sizeable fraction of the MBZ. The corresponding real space characteristic length scale can be estimated as:
\begin{eqnarray}
l_R \simeq  \frac{2 \pi }{ l_{\Omega_n}} \simeq   a  \sqrt{{2\pi  |\mathcal{C}_n| q} } ,
\end{eqnarray}
which is, for example, less than nine lattice spacings for the lowest bands with $\alpha =1/11$ and $ |\mathcal{C}_n| =1$. 

We note that as $\alpha \rightarrow 0$, the energy level spacing between the Harper-Hofstadter bands goes to zero and our single-band approximation is no longer valid. In this regime, the characteristic length scale is set instead by the interplay of the lattice and the harmonic trap, and can be estimated as  $l_R = \sqrt{ 1 / M \omega}$ where $M$ is the particle mass and $\omega$ is the harmonic trapping frequency.   

Similar results for $l_{\Omega_n}$ will hold for any system where the energy bands of $\mathcal{H}_0$ are characterised by Chern numbers and can be considered flat compared with the harmonic trapping energy. For such a model,  $l_{\Omega_n} \propto  p_{BZ} \sqrt{ 1 / 2 \pi \mathcal{C}_n}$,  if we approximate the BZ as a square area, $A_{BZ}$, of side $ p_{BZ}$. This is a sizeable fraction of the BZ even for large Chern numbers. For example, when $\mathcal{C}_n=15$, low-energy eigenstates are still delocalised over $\sim10\%$ of the BZ, or on the order of ten lattice sites in real space (assuming $ p_{BZ} = 2\pi / a$). 

If the bandwidth is not negligible, there are competing effects from the non-uniformity of the Berry curvature and of the energy band. In such systems, the eigenstates will not be Landau level-like, but may be analogous to other known magnetic eigenstates. Example systems of this type will be the subject of an upcoming publication. We note that the effect of both stronger Berry curvature or non-uniform energy band dispersions will generally be to confine the particle in momentum space, leading to more delocalisation in real space. However, the possible phenomenology is as rich as can be imagined for particles in non-uniform magnetic fields and external potentials.